\documentclass[prd,amssymb,a4paper]{revtex4}
\usepackage{latexsym}
\usepackage{amsfonts,amsbsy}

\def\onehalf{\textstyle{\frac{1}{2}}}
\def\ihalf{\textstyle{\frac{i}{2}}}
\def\GR{General Relativity}
\def\gammabol{{\stackrel{\circ}{\Gamma}}{}}
\def\omegabol{{\stackrel{\circ}{\omega}}{}}

\def\Rbol{{\stackrel{\circ}{R}}{}}

\begin{document}
\title{Gravitation as Anholonomy}
\author{R. Aldrovandi}
\email{ra@ift.unesp.br}

\author{P. B. Barros}

\author{J. G. Pereira}
\email{jpereira@ift.unesp.br}

\affiliation{Instituto de F\'{\i}sica Te\'orica,
Universidade Estadual Paulista \\
Rua Pamplona 145, 01405-900 S\~ao Paulo SP, Brazil}

\date{\today}

\begin{abstract}
A gravitational field can be seen as the anholonomy of the tetrad fields.
This is more explicit in the teleparallel approach, in which the gravitational
field-strength is the torsion of the ensuing Weitzenb\"ock connection.
In a tetrad frame, that torsion is just the anholonomy of that frame. 
The infinitely many tetrad fields taking the Lorentz metric into a given
Riemannian metric differ by point-dependent Lorentz transformations. Inertial
frames constitute a smaller infinity of them, differing by fixed-point  Lorentz
transformations. Holonomic tetrads take the Lorentz metric into itself,
and correspond to Minkowski flat spacetime. An accelerated frame 
is necessarily anholonomic and sees the electromagnetic
field strength with an additional term.
\end{abstract}
\maketitle


\section{Introduction}

Anholonomy --- the property of a differential form which is not the
differential of anything, or of a vector field which is not a gradient
--- is commonplace in many chapters of Physics. Heat and work, for
instance, are typical anholonomic coordinates on the space of
thermodynamic variables, and the angular velocity of a generic rigid
body is a classical example of anholonomic velocity. In gravitation
theory, however, anholonomy does not seem to have had its pervading
role as emphasized as it should. We intend here to fill in that gap,
by bringing to the forefront the anholonomic character of some
well-known features.

We are going to use the notation $\{e_{a}, e^{a}\}$ for general linear frames, and
$\{h_{a}, h^{a}\}$ for a generic tetrad field, with the Lorentz
indices $a, b, c, \dots = 0, 1, 2, 3$ raised and lowered by the Lorentz metric
\[
\eta = \eta^{-1} = {\rm diag}(1,-1,-1,-1).
\]
Greek indices $\mu, \nu, \rho, \dots = 0,1,2,3$  will refer to the Riemannian
spacetime. Curve parameters will be indicated by $u$ and $v$, with the
correspondent tangent fields denoted by the  respective capitals $U$ and $V$, as in
\[
U = \frac{d}{d u} = U^\lambda \partial_\lambda =
\frac{d x^\lambda}{d u}\ \partial_\lambda.
\]
The notation $i, j, k, \dots = 1,
2, 3$ is reserved for space indices. Parenthesis $(\mu \nu \rho \dots)$ and
brackets $[\mu \nu \rho \dots]$ indicate symmetrization and
antisymmetrization of included indices. Thus,
${\Gamma}{}^{\lambda}{}_{(\mu \nu)}$ =
$\onehalf ({\Gamma}{}^{\lambda}{}_{\mu \nu} +
{\Gamma}{}^{\lambda}{}_{\nu \mu})$ and
${\Gamma}{}^{\lambda}{}_{[\mu \nu]}$ = $\onehalf
({\Gamma}{}^{\lambda}{}_{\mu \nu} - {\Gamma}{}^{\lambda}{}_{\nu
\mu})$ designate the symmetric and antisymmetric parts of
${\Gamma}{}^{\lambda}{}_{\mu \nu}$.

A spacetime is a 4-dimensional Riemannian manifold whose tangent space 
at each point is a Minkowski spacetime~\cite{Syn60}. Consider, on such a general
spacetime, a coordinate system $\{x^\mu\}$, and  also a coordinate system 
$\{y^a\}$ on the tangent Minkowski spacetime. Such coordinate systems define, on
their domains of definition, local bases for vector fields, formed by the sets of
gradients
$\{\frac{\partial \ }{\partial x^\mu}\}$,
$\{\frac{\partial
\ }{\partial y^a}\}$, as well as bases $\{dx^\mu\}$, $\{dy^a\}$ for
covector fields, or differentials.  These bases are dual, in
the sense that $dx^\mu(\frac{\partial \ }{\partial x^\nu})$ =
$\delta^\mu_\nu$ and $dy^a(\frac{\partial \ }{\partial y^b})$ =
$\delta^a_b$. On the respective domains of definition, any vector or covector
can be expressed in terms of these bases, which can furthermore be extended by
direct product to constitute bases for general tensor fields.

A ``holonomic'' base like $\{\frac{\partial \
}{\partial x^\mu}\}$, related to coordinates, is a very particular case of
linear base.  Any set of four linearly independent fields
$\{e_{a}\}$  will form another base, and will have a dual
$\{e^{a}\}$ whose members are such that $e^{a}(e_b) =
\delta^a_b$.  These frame fields are the general linear bases on the spacetime
differentiable manifold whose set, under conditions making of it also a
differentiable manifold, constitutes the bundle of linear frames.  Of
course, on the common domains they are defined, the members of a base can be
written in terms of the members of the other: $e_a = e_a{}^\mu
\partial_\mu$, $e^{a} = e^{a}{}_\mu dx^\mu$, and conversely.  We can consider
general transformations taking any base $\{e_{a}\}$ into any other set
$\{e'_{a}\}$ of four linearly independent fields.  These
transformations constitute the linear group $GL(4, {\mathbb R})$ of
all real $4 \times 4$ invertible matrices.  Notice that these frames,
with their bundle, are constitutive parts of spacetime.  They are
automatically present as soon as spacetime is taken to be a
differentiable manifold~\cite{KN96}. 

Consider the metric $g$ which has components $g_{\mu \nu}$ in some
dual holonomic base $\{d x^{\mu}\}$:
\begin{equation}
	 g = g_{\mu \nu} dx^{\mu} \otimes dx^{\nu} = g_{\mu \nu} dx^{\mu}
	 dx^{\nu}.  \label{eq:Riemetric}
\end{equation}
A tetrad field $\{h_{a} =
h_{a}{}^{\mu} \frac{\partial}{\partial x^{\mu}}\}$ will be a  linear base which
relates $g$ to the Lorentz metric $\eta =
\eta_{a b} dy^a dy^b$ by
\begin{equation}
 \eta_{a b} = g(h_{a},h_{b}) = g_{\mu \nu} h_{a}{}^{\mu}
 h_{b}{}^{\nu}.  \label{eq:gtoeta}
\end{equation}
This means that a tetrad field is a linear frame whose members $h_{a}$ are
(pseudo-)orthogonal by the metric $g$.  We shall see later how two of such 
bases are related by the Lorentz subgroup of  the linear group
$GL(4, {\mathbb R})$. The components of the dual base members
$\{h^{a} = h^{a}{}_{\nu} dx^{\nu} \}$ satisfy
\begin{equation}
h^{a}{}_{\mu} h_{a}{}^{\nu} = \delta_{\mu}^{\nu}\ \ {\rm and} \ \
h^{a}{}_{\mu} h_{b}{}^{\mu} = \delta^{a}_{b}, \label{eq:tetradprops1}
\end{equation}
so that Eq.~(\ref{eq:gtoeta}) has the converse
\begin{equation}
g_{\mu \nu} = \eta_{a b}\ h^{a}{}_{\mu} h^{b}{}_{\nu}.
    \label{eq:tettomet}
\end{equation}
We shall be almost exclusively interested in tetrad fields. In consequence, though
many of our later statements --- such as those given in Eqs.
(\ref{eq:comtable}-\ref{eq:omegagain}) below ---  hold for general linear frames,
we shall specialize them accordingly.

An important point we would like to stress is that anholonomy is related to
the very existence of a gravitational field. Given a Riemannian metric as in
(\ref{eq:tettomet}), the presence or absence of a gravitational field is fixed by
the anholonomic or holonomic character of the forms $h^{a} = h^{a}{}_{\nu}
dx^{\nu}$. We can think of a change of coordinates $\{y^a\} \Leftrightarrow
\{x^\mu\}$ represented by
\[
dy^a = \frac{\partial y^a}{\partial x^\mu}\ dx^\mu = dy^a \left(\frac{\partial
\ }{\partial x^\mu} \right) \, dx^\mu.
\]
The 1-form $dy^a$ is holonomic, just the differential of the coordinate
$y^a$, and the objects $\{\frac{\partial y^a}{\partial x^\mu}\}$ are
the components of the holonomic form $dy^a$ written in the base $\{dx^\mu\}$.
Thus, such a coordinate change is just a change of holonomic bases of
1-forms.

Take now a dual base $\{h^a\}$ such that $d h^a \ne 0$, which is not
formed by  differentials. Apply the anholonomic 1-forms $h^a$ (such that $d h^a
\ne 0$) to $\frac{\partial \ }{\partial x^\mu}$. The results, $h^a{}_\mu =
h^a(\frac{\partial \ }{\partial x^\mu})$, give the components of each $h^a$ =
$h^a{}_\mu dx^\mu$ along $dx^\mu$. The procedure can be inverted when the
$h^a$'s are linearly independent, and defines vector fields $h_a$ =
$h_a{}^\mu \frac{\partial \ }{\partial x^\mu}$ which are not gradients. Because
closed forms are locally exact, holonomy/anholonomy can be given a trivial
criterion: A form is holonomic {\it iff\ } its exterior derivative vanishes.  A
holonomic tetrad will always be of the form $\{h^{a} = d y^a\}$ for some
coordinate  set $\{y^a\}$. For such a tetrad, the metric tensor
(\ref{eq:tettomet}) would be simply the components of the Lorentz metric
$\eta$ transformed to the coordinate system $\{x^\mu\}$. The Levi-Civita
connection, or Christoffel symbol,
\begin{equation} %
\gammabol{}^{\lambda}{}_{\mu \nu} = {\textstyle
\frac{1}{2}} g^{\lambda \rho} \left[\partial_{\mu} g_{\rho \nu} +
\partial_{\nu} g_{\rho \mu} - \partial_{\rho} g_{\mu \nu} \right],
\label{Christoffel}
\end{equation}
leads to a Riemann curvature tensor --- the gravitational field
strength in \GR\  --- which vanishes if $\{h_a\}$ is  holonomic. A
gravitational field is present only when the tetrad fields are anholonomic.

Teleparallelism~\cite{hs} provides an approach to gravitation which is both
alternative and equivalent to \GR. The teleparallel presentation
of gravity is closer to the gauge-theoretical paradigm~\cite{hehl}
and thereby stresses the similarities between gravitation and the other
fundamental interactions~\cite{ap1}. It stresses also their main difference:
By putting the accent on the tetrad frames, it highlights the inertial
character of the gravitational force. In teleparallel gravity, the 
Weitzenb\"ock connection
\begin{equation} %
{\bar \Gamma}{}^{\lambda}{}_{\mu \nu } = h_a{}^{\lambda}
\partial_{\nu} h^{a}{}_{\mu} 
\label{eq:weitzen}
\end{equation} %
plays a central part: Its torsion will be the gravitational field strength. 
We shall for that reason pay special attention to the torsions of linear
connections. It should be remarked that for holonomic tetrads ${\bar \Gamma}$ is
torsionless.

Our policy will be to review well-known facts while emphasizing their
anholonomic content.  After some preliminaries on connections and
their torsions in section \ref{sec:linconn}, we proceed to a {\it
resum\'e} on three metric-related structures: The tetrad
fields, the Levi-Civita connection, and the Weitzenb\"ock connection. In
section \ref{sec:tetrads} we review the usual lore on tetrad fields as
introduced through the metric they determine, and section \ref{sec:Chris} is
devoted to the Levi-Civita connection. Non-inertial frames are discussed in
section \ref{sec:nonin}, in which it is shown that accelerated frames are
necessarily anholonomic. A synopsis on teleparallelism is given in section
\ref{sec:telepar}. The last section sums it all up and
adds some comments on remaining questions.

\section{Linear Connections}
\label{sec:linconn}

Linear connections have a great degree of intimacy with spacetime
because they are defined on the bundle of linear frames, which is a
constitutive part of its manifold structure. That bundle has some properties not
found in the bundles related to gauge theories~\cite{AP95b}.  Mainly, it exhibits
soldering, which leads to the existence of torsion for every
connection~\cite{KN96}.  Linear connections --- in particular, Lorentz
connections --- always have torsion, while gauge potentials have not. 
The torsion
$T$ of a linear connection $\Gamma$ in a linear frame is just the covariant derivative of the frame
members. In a  holonomic base, the torsion components are
essentially the antisymmetric parts of the connection components:
\begin{equation}
    T^{\lambda}{}_{\mu \nu} = \Gamma^{\lambda}{}_{\nu \mu} -
    \Gamma^{\lambda}{}_{\mu \nu} = -\ 2\ \Gamma^{\lambda}{}_{[\mu
    \nu]}.
    \label{eq:torsion}
\end{equation}
When $T^{\lambda}{}_{\mu \nu} \ne 0$ it will be impossible to make all
the components $\Gamma^{\lambda}{}_{\mu \nu}$ equal to zero in a
holonomic base. Torsion has important consequences, even if vanishing: The
property
$T^{\lambda}{}_{\mu
\nu} = 0$, which holds for the Levi-Civita connection of a metric, is at
the origin of the well-known cyclic symmetry of the Riemann tensor
components.

The condition of metric compatibility is that the metric be
everywhere parallel-transported by the connection, that is,
$
\nabla_\lambda\ g_{\mu \nu} \equiv \partial_\lambda g_{\mu \nu} -
\Gamma^\rho{}_{\mu \lambda} g_{\rho \nu} - \Gamma^\rho{}_{ \nu
\lambda} g_{\mu \rho} = 0,
$
or equivalently
\begin{equation}
	\partial_\lambda g_{\mu \nu} = 2\ \Gamma_{(\mu \nu) \lambda},
\label{compatibility}
\end{equation}
where we have used the notation $\Gamma_{\mu \nu \lambda} = g_{\mu \rho} \,
\Gamma^\rho{}_{\nu \lambda}$. A metric defines a Levi-Civita connection
$\gammabol{}$, which is that unique connection which satisfies this
condition and has zero torsion.  Its components in a holonomic base are
the Christoffel symbols (\ref{Christoffel}).
If a connection $\Gamma$ preserves a metric and is not its Levi-Civita
connection, then it will have $T^{\lambda}{}_{\mu \nu} \ne 0$.  The difference
between two connections is a tensor.  The expression
\begin{equation}
K{}^{\lambda}{}_{\mu \nu } = \gammabol{}^{\lambda}{}_{\mu \nu } -
\Gamma^{\lambda}{}_{\mu \nu } \label{decomp1}
\end{equation}
defines the contorsion tensor $K$ of $\Gamma$.  Using
(\ref{compatibility}) both for $\Gamma$ and $\gammabol{}$, we have
$\Gamma_{(\lambda \mu ) \nu}$ = $\gammabol_{(\lambda \mu ) \nu}$ and
consequently
\begin{equation}
K_{(\lambda \mu ) \nu} = 0. \label{Ktraceless}
\end{equation}
Metric compatibility gives one further constraint: Contorsion is fixed
by the torsion tensor:
\begin{equation}%
K^{\lambda}{}_{\mu \nu} = {\textstyle \frac{1}{2}}
\left[T^{\lambda}{}_{\mu \nu} + T_{\mu \nu }{}^{\lambda} + T_{
\nu \mu}{}^{\lambda}\right].  \label{contorsion}
\end{equation}%
As both $T$ and $K$ are tensors, this relationship holds in any basis. 

When we say that some field (vector, covector, tensor, spinor) is {\em
everywhere} parallel-transported by a connection, we mean the
vanishing of the corresponding covariant derivative all over the
domain on which field and connection are defined.  This is a very
strong condition.  Most frequently, the interest lies in
parallel-transport along a curve.  Thus, for example, the geodesic
equation
\begin{equation}
    \frac{\nabla U^{\lambda}}{\nabla u} \equiv \frac{dU^{\lambda}}{du}  +
\Gamma^{\lambda}{}_{\mu \nu} U^{\mu}
    U^{\nu} = 0
    \label{eq:geodesic1}
\end{equation}
defines a curve $\gamma(u)$ whose velocity field $U$ itself is
parallel-transported by $\Gamma$ along the curve.  For a general connection, 
this equation defines a  {\em self-parallel curve}. 
Each connection defines an acceleration which is given by the so-called
equation of force
\begin{equation}
 \frac{\nabla U^{\lambda}}{\nabla u} = a^\lambda.
\label{eq:force}
\end{equation}

\section{The Class of Frame Fields of a Metric}
\label{sec:tetrads}

The base $\{h_{a}\}$ is far from being unique.  There exists actually a six-fold
infinity of tetrad fields $\{h_{a} = h_{a}{}^{\mu} \frac{\partial}{\partial
x^{\mu}}\}$, each one relating $g$ to the Lorentz metric $\eta$ by  Eqs.
(\ref{eq:gtoeta}-\ref{eq:tettomet}). This comes from the fact that, at each point
of the Riemannian spacetime, Eq.~(\ref{eq:tettomet}) only determines the tetrad
field up to transformations of the six-parameter Lorentz group in the anholonomic
indices. Suppose in effect another tetrad $\{h'_{a}\}$ such that
\begin{equation}
g_{\mu \nu} = \eta_{a b}\ 
h^{a}{}_{\mu} h^{b}{}_{\nu} = \eta_{c d}\ 
h^{' c}{}_{\mu} h^{' d}{}_{\nu}. \label{etatogmunu}
\end{equation}
Contracting both sides with $h_{e}{}^{\mu}h_{f}{}^{\nu}$, we arrive at
$$%
\eta_{a b} = \eta_{c d}\ 
(h^{' c}{}_{\mu} h_{a}{}^{\mu} ) (h^{' d}{}_{\nu} h_{b}{}^{\nu}).
$$%
This equation says that the matrix with entries 
\begin{equation}
\Lambda^a{}_{b} = h^{' a}{}_{\mu}\ h_{b}{}^{\mu}, \label{Lortetrad}
\end{equation}
which gives the transformation 
\begin{equation}
 h^{' a}{}_{\mu} = \Lambda^a{}_{b}\  h^{b}{}_{\mu},
\end{equation}
satisfies
\begin{equation}
 \eta_{cd} \ \Lambda^c{}_{a}\ \Lambda^d{}_{b} = \eta_{a b}.
\end{equation}
This is just the condition that a matrix $\Lambda$ must satisfy in
order to belong to (the vector representation of) the Lorentz group.

Basis $\{h_{a}\}$ will be anholonomic --- unrelated to any coordinate
system --- in the generic case.  This means that, given the
commutation table
\begin{equation}
    [h_{a}, h_{b}] = f^{c}{}_{a b}\ h_{c},
    \label{eq:comtable}
\end{equation}
there will be non-van\-ishing structure coefficients $f^{c}{}_{a b}$ for some
$a, b, c$.  The frame $\{\frac{\partial\ }{\partial x^{\mu}}\}$ has been
presented above as holonomic precisely because its members commute with each
other. The dual expression of the commutation table above is the Cartan
structure equation
\begin{equation}
    d h^{c} = -\ \onehalf f^{c}{}_{a b}\ h^{a} \wedge h^{b} =   \onehalf \ 
(\partial_\mu h^c{}_\nu - \partial_\nu h^c{}_\mu)\ dx^\mu \wedge dx^\nu.
    \label{eq:dualcomtable}
\end{equation}
The structure coefficients represent the curls of the base members:
\begin{equation}
f^c{}_{a b}  = h^c ([h_a, h_b]) = h_a{}^{\mu} h_b{}^{\nu} (\partial_\nu
h^c{}_{\mu} - 
\partial_\mu h^c{}_{\nu} ) =  h^c{}_{\mu} [h_a(h_b{}^{\mu}) - 
h_b(h_a{}^{\mu})]. \label{fcab}
\end{equation}
If $f^{c}{}_{a b}$ = $0$, then $d h^{a} = 0$ implies the local
existence of functions (coordinates) $y^a$ such that $h^{a}$ = $d
y^a$. The tetrads are gradients when the curls vanish. 

Equation (\ref{eq:tettomet}) tells us that the components of metric
$g$, in the tetrad frame, are just those of the Lorentz metric. This does {\it
not} mean that the frame is inertial, because the metric derivatives --- which
turn up in the expressions of forces and accelerations --- are not
tensorial.  In order to define derivatives with a well-defined tensor
behavior (that is, which are covariant), it is essential to introduce
connections $\Gamma^\lambda{}_{\mu \nu}$, which are vectors in the
last index but whose non-tensorial behavior in the first two indices
compensates the non-tensoriality of the usual derivatives. 
Connections obey in consequence a special law: In the tetrad
frame, a connection $\Gamma$ has components
\begin{equation}
\omega^{a}{}_{b c} = h^{a}{}_{\lambda} \left[h_{c} (h_{b}{}^{\lambda})
+ h^{a}{}_{\lambda}\ {\Gamma}^{\lambda}{}_{\mu \nu}\ h_{b}{}^{\mu}
h_{c}{}^{\nu} \right] \equiv h^{a}{}_{\lambda} \nabla_{c}
(h_{b}{}^{\lambda})
\label{eq:omegagain}.
\end{equation}
This transformation law ensures the tensorial behavior of the
covariant deri\-vative: $\nabla_\nu V^\lambda = h^{a}{}_{\nu}
h_b{}^\lambda \nabla_a V^b = h^{\prime a}{}_{\nu} h^\prime_b{}^\lambda
\nabla^\prime_a V^{\prime b}$ and $\nabla_a V^b = \Lambda^c{}_a
\Lambda^b{}_d \nabla^\prime_c V^{\prime d}$.  The antisymmetric part
 of $\omega^{a}{}_{b c}$ in the last two indices can be computed by
using Eqs.~(\ref{eq:torsion}) and (\ref{fcab}).  The result shows that
torsion, seen from the anholonomic frame, includes the anholonomy:
\begin{equation}
T^{a}{}_{b c}  = -\ f^{a}{}_{b c} - (\omega^{a}{}_{b c} -
\omega^{a}{}_{c b}). \label{torsionfrom}
\end{equation}

There is a constraint on the first two indices of $\omega^{a}{}_{b c}$
{\em if}\ \ $\Gamma$ preserves the metric. In effect, Eqs. 
(\ref{compatibility}) and (\ref{eq:gtoeta}) lead to
\begin{equation}%
\omega_{a b c} = - \ \omega_{b a c}.
\label{omisLor1}
\end{equation}%
This antisymmetry in the first two indices, after lowering with the
Lorentz metric, says that $\omega$ is a Lorentz connection. This is to say that
it is of the form
\[
\omega = \onehalf\ J_a{}^b \, \omega^a{}_{bc} \, h^c,
\]
with $J_a{}^b$ the Lorentz generators written in an appropriate representation.
Therefore, any connection preserving the metric appears, when
its components are written in the tetrad frame, as a Lorentz-algebra valued
1-form. If we use Eq.~(\ref{Lortetrad}) and the inverse
$(\Lambda^{-1})^{a}{}_b = h^{a}{}_{\mu} \, h'_{b}{}^{\mu}$ = $\eta_{bc}
\, \eta^{ad} \, \Lambda^{c}{}_{d}$ = $\Lambda_b{}^a$, we find how the components
change under tetrad (Lorentz) transformation:
\begin{equation}
\omega^{'a}{}_{b \nu} = \Lambda^a{}_c\ \omega^{c}{}_{d \nu}
(\Lambda^{-1})^{d}{}_b + \Lambda^a{}_c\ \partial_\nu
(\Lambda^{-1})^{c}{}_b. \label{omisLor}
\end{equation}
This establishes the connection $\omega$ (which is $\Gamma$ with
components written in any tetrad frame) as a Lorentz connection. For such Lorentz
connections, use of (\ref{torsionfrom}) for three combinations of the indices
gives
\begin{equation}%
\omega^{a}{}_{b c}  = -\ \onehalf\ (f^{a}{}_{b c} + T^{a}{}_{b c} +
f_{b c}{}^{a} + T_{b c}{}^{a} + f_{c b}{}^{a} + T_{c b}{}^{a}).
\label{tobetaken2}
\end{equation}%

The components of a velocity $U$ are given by the holonomic form $dx^\mu$ applied
to the time-evolution vector field $\frac{d\ }{du}$, that is,
\[
U^\mu = \frac{dx^\mu}{du} = dx^\mu \left(\frac{d\ }{du} \right).
\]
The velocity $U^\mu$ represents, consequently, the variation of the coordinate
$x^\mu$ in time $u$. In the tetrad frame $\{h_a\}$, $U$ has components 
\begin{equation}
U^a = h^a{}_\mu U^\mu = h^a{}_\mu dx^\mu
\left(\frac{d\ }{du} \right) = h^a \left(\frac{d\ }{du} \right).
\label{Ua}
\end{equation}
If $\{h_a\}$ is holonomic, then $h^a$ = $\frac{\partial y^a}{\partial
x^\mu}dx^\mu$ for some coordinates $\{y^a\}$, and $U^a$ measures the
variation of coordinate $y^a$ in time $u$. If $\{h_a\}$ is
not holonomic, however, $U^a$ will be an anholonomic velocity: Its
components will be the variations of {\it no} coordinates with time (a
classical non-relativistic example has been mentioned in the Introduction, the
angular velocity of a rigid body in the general, non-planar case).  We
have said that the tetrad frame ``sees'' everything in terms of the flat,
Minkowski space coordinates. The difference with respect to ``native''
special-relativistic objects lies in the anholonomic character of the
frame. An usual holonomic velocity $U^\mu$ in Riemann spacetime, for example,
becomes, in the tetrad frame, an anholonomic velocity, whose
components $U^a$ in flat Minkowski space are not derivatives of any coordinate
with respect to time. A ``native'' special-relativistic observer would see a
holonomic velocity $V^a = d x^a/d\sigma$, with $d\sigma^2 =
\eta_{a b} \ dx^a dx^b$. In the tetrad frame $\{h_{a}\}$, the equation of
force (\ref{eq:force}) has the form
\begin{equation}
    \frac{d U^{a}}{du} + \omega^{a}{}_{b c}\ U^{b}
U^{c} = a^a,
    \label{eq:force2}
\end{equation}
where $\omega^{a}{}_{b c}$ and $U^a$ are given by (\ref{eq:omegagain}) and
(\ref{Ua}) respectively.

The Riemannian metric $g = (g_{\mu \nu})$ is a Lorentz invariant, 
for which any two tetrad fields as $\{h_{a}\}$ and $\{h'_{a}\}$ in
(\ref{etatogmunu}) are equivalent.  A metric corresponds to an equivalence class
of tetrad fields, the quotient of the set of all tetrads by the Lorentz group. 
The sixteen fields $h^{a}{}_{\mu}$ correspond, from the field-theoretical point
of view, to ten degrees of freedom --- like the metric --- once the equivalence
under the six-parameter Lorentz group is taken into account.

The tetrads belong to the carrier space of a matrix representation of
the Lorentz group. They have, however, a very special characteristic:
They are themselves invertible matrices. A group element taking some
member of the representation space into another can in consequence be
written in terms the initial and final members, as in (\ref{Lortetrad}). 
This establishes a deep difference with respect to the other fundamental
interactions, described by gauge theories.  There are matrix representations
in gauge theories, like the adjoint representation, but their members are not
invertible.

\section{A Preferred Connection}
\label{sec:Chris}

A metric $g$ defines a preferred connection, the Levi-Civita
connection $\gammabol$ given by (\ref{Christoffel}) which is, we repeat, the
single connection preserving $g$ which has zero torsion. Its curvature
Riemann tensor,
\[
\Rbol^\lambda{}_{\rho \mu\nu} = \partial_\mu \gammabol^{\lambda}{}_{\rho \nu} 
- \partial_\nu \gammabol^{\lambda}{}_{\rho \mu} +
\gammabol^{\lambda}{}_{\sigma \mu} \gammabol^{\sigma}{}_{\rho \nu} -
\gammabol^{\lambda}{}_{\sigma \nu} \gammabol^{\sigma}{}_{\rho \mu},
\]
is the covariant representative of the gravitational field in General
Relativity. The Lorentz connection $\omegabol$ obtained via a
tetrad field $h_{a}$ is, in this case, usually called ``spin-connection''. 
It appears, for example, in the Dirac equation~\cite{dirac}
\begin{equation} %
i \hbar \gamma^c h_c{}^\mu \left( \partial_\mu - \frac{i}{4} \,
\omegabol{}^{a b}{}_{\mu} \, \sigma_{a b} \right) \psi \equiv i \hbar
\gamma^c \left( h_c - \frac{i}{4} \, \omegabol{}^{a b}{}_{c} \,
\sigma_{a b} \right) \psi = m c \psi,
\label{cde}
\end{equation} %
with $\sigma_{a b} = \ihalf [\gamma_a, \gamma_b]$ the spinor representation of the
Lorentz generators. Its components are related to $\gammabol^{\lambda}{}_{\mu \nu}$
by
\begin{equation} %
    \omegabol^{a}{}_{b \nu} = h^{a}{}_{\lambda}\
    \gammabol^{\lambda}{}_{\mu \nu}\ h_{b}{}^{\mu} +
    h^{a}{}_{\rho} \ \partial_{\nu} h_{b}{}^{\rho}. 
    \label{eq:gamtomegabol} 
\end{equation} %
This expression, combined with (\ref{fcab}), gives
\begin{equation} %
	\omegabol^{a}{}_{b c} - \omegabol^{a}{}_{c b} = f^{a}{}_{c b}.
\label{omandf}
\end{equation} %
We see that, once looked at from the frame $\{h_{a}\}$, the symmetric
connection $\gammabol$ acquires an antisymmetric part, which has only to do with
the anholonomy of the basis. That this is a mere artifact due to the frame
anholonomy is better seen in an example in electromagnetism.  In effect, a
symmetric connection does not alter the expression $F_{\mu \nu}$ =
$\partial_{\mu} A_{\nu} - \partial_{\nu} A_{\mu}$ of the field
strength in terms of the electromagnetic potential $A_{\mu}$. In frame
$\{h_{a}\}$, however,
\[%
h_{a}{}^{\mu} h_{b}{}^{\nu}\left( \partial_{\mu} A_{\nu} - 
\gammabol^{\lambda}{}_{\nu \mu} A_{\lambda} \right) = 
h_{a} A_{b} - \omegabol^{c}{}_{ba} \, A_{c}, 
\]%
so that (\ref{omandf}) leads to
\begin{equation} %
F_{ab} = h_{a} (A_{b}) - h_{b} (A_{a}) - f^{c}{}_{ab}  A_{c}.
\label{Finh}
\end{equation} %
On the other hand, this is exactly what comes out from a direct calculation of
the invariant form $F = dA$ = $d(A_{a} h^{a})$ by using
(\ref{eq:dualcomtable}) {\em in the absence of any connection}. Notice that
the last term in the expression above is essential to the invariance of
$F_{ab}$ under a $U(1)$ gauge transformation as seen from the frame
$\{h_{a}\}$, which is $A_{a} \rightarrow A'_{a} = A_{a} + h_{a} \phi$:
$$%
F'_{ab} = h_{a} A'_{b} - h_{b} A'_{a} + f^{c}{}_{ab}
A'_{c} = F_{ab} + h_{a} h_{b} \phi - h_{b} h_{a} \phi -
f^{c}{}_{ab} h_{c} \phi = F_{ab}.
$$%

The force equation (\ref{eq:force2})  can be expressed, by using
(\ref{tobetaken2}) with $T^a{}_{b c} = 0$, in terms of the anholonomy
coefficients as
\begin{equation}
   \frac{d U^{a}}{du} +
 f_{b}{}^a{}_c\ U^{b} U^{c} = a^a.
\end{equation} %
The Riemann curvature tensor will have tetrad components
\[
\Rbol^{a}{}_{b c d} = h^{a}{}_{\rho} h_{b}{}^{\sigma} h_{c}{}^{\mu}
h_{d}{}^{\nu} \, \Rbol^{\rho}{}_{\sigma \mu \nu},
\]
which gives
\begin{equation} %
\Rbol^{a}{}_{b c d} = h_{c}\ \omegabol^{a}{}_{b d} - h_{d}\
\omegabol^{a}{}_{b c} + \omegabol^{a}{}_{e c} \, \omegabol^{e}{}_{b d} -
\omegabol^{a}{}_{e d} \, \omegabol^{e}{}_{b c} - f^e{}_{c d} \,
\omegabol^{a}{}_{b e}.
\label{riemm}
\end{equation} %

\section{Non-Inertial Frames}
\label{sec:nonin}

Another tetrad frame $\{h'_a\}$ will see another spin connection, that
is, will see the connection $\omegabol$ with other components, as given by
(\ref{omisLor}). Suppose for a moment the frame $\{h'_a\}$ to be such that
$\omegabol'^{a}{}_{b \nu} = 0$ (such a frame does exist at each point,
and along a differentiable curve, see below). 
In that case $\omegabol$ would be a pure gauge of the Lorentz group,
\begin{equation}
\omegabol^{a}{}_{b \nu} = h^c{}_\nu\ \omegabol^{a}{}_{b c}=
(\Lambda^{-1})^a{}_c \partial_\nu \Lambda^{c}{}_b = (\partial_\nu \ln
\Lambda)^a{}_b.  \label{omegazerox}
\end{equation}
Matrix $\Lambda$ has
the form
\[
\Lambda = \exp{W} =
\exp \left[{\onehalf \, J_{c d} \, \alpha^{c d}} \right],
\]
with $J_{c d}$ denoting the generators and $\alpha^{c d}$ the parameters of the
Lorentz transformation. Therefore,
\begin{equation}
\omegabol^{a}{}_{b}  = (\Lambda^{-1} d \Lambda)^{a}{}_{b}  =
(d \ln \Lambda)^{a}{}_{b} = (d W)^{a}{}_{b} = d W^{a}{}_{b}.
\end{equation}
Furthermore, using the vector representation for $J_{c d}$,
\[%
W^{a}{}_{b} = \onehalf (J_{c d})^a{}_b \ \alpha^{c d} = \onehalf
(\eta_{db}\ \delta^a_c - \eta_{cb}\ \delta^a_d) \ \alpha^{c d} =
\onehalf (\alpha^{a}{}_{b} -\alpha_{b}{}^{a}) = \alpha^{a}{}_{b},
\]%
so that Eq.~(\ref{omegazerox}) is the same as
\begin{equation}
\omegabol^{a}{}_{b c} = h_c (\alpha^{a}{}_{b}). \label{omegazero2}
\end{equation}
Notice that $\Lambda$ represents here that very special Lorentz
transformation taking $\{h_a\}$ into a tetrad  $\{h'_a\}$ in which the
connection has vanishing components. From Eq.~(\ref{eq:omegagain}) written for
$h'_{a}$ in the form
$$%
\partial_{\nu} \ h'_{a}{}^{\lambda} + \gammabol^{\lambda}{}_{\mu \nu}\
h'_{a}{}^{\mu} = h'_{c}{}^{\lambda}\ \omegabol^{\prime c}{}_{a \nu},
$$%
the condition $\omegabol^{\prime c}{}_{a \nu} = 0$, if valid on a general
domain, would lead to vanishing curvature.  Take however the integral
curve $\gamma$ of a vector field $U$ with fixed initial values.  Then,
the condition
\begin{equation}
U^\nu \partial_{\nu} \ h'_{a}{}^{\lambda} + \gammabol^{\lambda}{}_{\mu
\nu}\ h'_{a}{}^{\mu} U^\nu = h'_{c}{}^{\lambda}\ \omegabol^{\prime
c}{}_{a \nu} U^\nu = 0
\end{equation}
is possible even in the presence of curvature: It means simply that
the four tetrad vectors $h'_a$ are parallel-transported along $\gamma$. 
It is a deep result~\cite{Ili98,Iliev,Har95,Ili98b} that the connection
$\omegabol^{\prime c}{}_{a \nu}$ can be made to vanish at a point of $\gamma$ by
a choice of $\{h'_a\}$, and that this frame can be propagated along it while
preserving this property.  Each vector $h'_a$ will then feel no force along
$\gamma$, as $\omegabol^{\prime c}{}_{a \nu} U^\nu = 0$ all along.  This
characterizes an inertial frame, in which Special Relativity applies.  If the
curve is timelike, an observer attached to this frame will be an inertial
observer~\cite{HE73,Ell90,EE98}. As every other frame can be got from it at each
point by a Lorentz transformation, General Relativity appears as a gauge theory
for the Lorentz group {\it along the curve}.  Distinct curves require different
frames, and one same frame cannot be parallel-transported along two distinct
intersecting curves unless the Riemann curvature tensor vanishes. A
clear statement of the equivalence principle along these lines can
be found in Ref.~\cite{ABP03a}.

The timelike member $h_{0}$ of a set $\{h_{a}\}$ of vector fields constituting a
tetrad will define, for each set of initial conditions, an integral curve
$\gamma$. It is always possible to identify $h_{0}$ to the velocity
$U$ of $\gamma$. This would mean $U^a = h^a{}_\nu h_0{}^\nu = \delta^a_0$. The
frame, as it is carried along that timelike curve, will be inertial or not,
according to the corresponding force law.  The force equation can be obtained by
using, for example, Eq.~(\ref{eq:gamtomegabol}) written for $h_{0}$:
$$%
\partial_{\nu} \, h_{0}{}^{\lambda} + \gammabol^{\lambda}{}_{\mu \nu} \,
h_{0}{}^{\mu} = h_{a}{}^{\lambda}\ \omegabol^{a}{}_{0 \nu}.
$$%
This leads, with $U = \frac{d\ }{du} = h_{0}$, to the expression
$$ %
h_{0}{}^{\nu} \partial_{\nu} \ h_{0}{}^{\lambda} +
\gammabol^{\lambda}{}_{\mu \nu}\ h_{0}{}^{\mu} h_{0}{}^{\nu} = 
U^\nu \partial_{\nu} \ U^{\lambda} + \gammabol^{\lambda}{}_{\mu \nu}\ U^{\mu}
U^{\nu}  = h_{a}{}^{\lambda}\
\omegabol^{a}{}_{0 \nu} h_{0}{}^{\nu},
$$ %
implying the frame acceleration
\begin{equation} %
a{}^{\lambda} =
h_{a}{}^{\lambda}\ \omegabol^{a}{}_{0 0}.
    \label{accel1} 
\end{equation} %
The relation to anholonomy is given by Eq.~(\ref{tobetaken2}), torsion turning
up as an accelerating factor:
\begin{equation} %
a{}^{\lambda}  = h_{a}{}^{\lambda}\
\omegabol^{a}{}_{0 0} = -\  h_{a}{}^{\lambda}\ (
f_{00}{}^{a} + T_{00}{}^{a}) = -\ \eta_{0 c} h^{b \lambda}\ 
(f^c{}_{0 b} + T^c{}_{0 b}). 
    \label{accel2} 
\end{equation} %

 Let us examine what happens in the absence of torsion.  The
 acceleration is then measured by the timelike component of the tetrad
 commutators involving the timelike member,
\begin{equation} %
a{}^{\lambda}  = h_{k}{}^{\lambda} 
\omegabol^{k}{}_{0 0}  = h_{k}{}^{\lambda} f^0{}_{0 k} = h_{k}{}^{\lambda} d
h^0 \, ([h_{0}, h_{k}]). 
    \label{accel4} 
\end{equation} %
It follows that an accelerated frame is necessarily anholonomic: It must
have {\em at least} $f^0{}_{0 k} \ne 0$.  From Eq.~(\ref{omegazero2}), the
transformation to an inertial frame involves only time-derivatives of
boost parameters (essentially the relative velocity):
\begin{equation}
\omegabol^{k}{}_{00} = h_0 (\alpha^{k}{}_{0}). \label{omegazero3}
\end{equation}
In the inertial frame $h'$, the velocity of frame $h$ will
have for components the boost transformations: $U^{\prime c} = h'{}^c{}_\mu
h_0{}^\mu =
\Lambda^c{}_0$. Something about the behavior of the spacelike members of the
tetrad along the curve $\gamma$ can be obtained from Eq.~(\ref{eq:gamtomegabol})
for $h_{i}$. Indicating by $a_{(i)}{}^{\lambda}$ the covariant change rate of
$h_{i}{}^{\lambda}$, we find
\[
a_{(i)}{}^{\lambda} = \nabla_U h_{i}{}^{\lambda} =
h_{a}{}^{\lambda}\ \omegabol^{a}{}_{i 0} =
h_{a}{}^{\lambda} h_{i} (\alpha^{k}{}_{0}) = \onehalf h^{c \lambda}
(f_{ic0} + f_{0ci} + f_{c0i}) = h_{c}{}^{\lambda} h_0 (\alpha^c{}_i).
\]
 As
$$%
\nabla_U h_a{}^\lambda = h_c{}^\lambda\  \omegabol^c{}_{a \nu} U^\nu
$$%
for any $U$, the Fermi-Walker derivative will be
$$%
 \nabla_U^{(FW)} h_a{}^\lambda = \nabla_U h_a{}^\lambda + a_a
U^\lambda - U_a a^\lambda.
$$%
The particular case
$$ %
\nabla_U^{(FW)} h_0{}^\lambda = \nabla_U h_0{}^\lambda - U_0 a^\lambda = 0 
$$ %
implies that $h_0$ is kept tangent along the curve.  The other tetrad members,
however, rotate with angular velocity 
\begin{equation}
\omega^k = \onehalf\ \epsilon^{kij} \omegabol_{ij0} = 
\onehalf\ \epsilon^{kij} f^{i}{}_{j 0}, \label{Riccirot0}
\end{equation}
which shows the $\omegabol^a{}_{b c}$'s in their role of Ricci's
coefficient of rotation~\cite{fock}. 

As another example, by Eq.~(\ref{Finh}) the electromagnetic field,
when looked at from a non-inertial frame, will forcibly include
extra, anholonomy-related, terms:
\[
F_{0k} =  h_{0} (A_{k}) - h_{k} (A_{0}) - f^{a}{}_{0 k}
A_{a} = h_{0} (A_{k}) - h_{k} (A_{0}) + a_{k}
A_{0} + \epsilon_{k i j} \omega^{i} A^{j}
\]%
\[%
F_{jk} = h_{j} (A_{k}) - h_{k} (A_{j}) - f^{a}{}_{j k} \, A_{a}. 
\]%
In the simplest gauge ($A^0$ = 0), the electric field reduces to the
Euler derivative
\begin{equation} %
 {\bf E} = \frac{d {\bf A}}{du}
 + {\boldsymbol \omega} \wedge {\bf A}.
\end{equation} %

\section{Teleparallelism}
\label{sec:telepar}

Each tetrad $\{h_{a}\}$ defines a special connection,
the Weitzen\-b\"ock connection given by (\ref{eq:weitzen}).  That
connection has some very interesting properties:\vspace{-6pt}
\begin{enumerate}
    
       \item It has vanishing components in the tetrad frame
 $\{h_{a}\}$ itself~\cite{hs}:
    \begin{equation} %
    {\bar \omega}^{a}{}_{b \nu} = h^{a}{}_{\lambda}\,
    {\bar \Gamma}^{\lambda}{}_{\mu \nu}\, h_{b}{}^{\mu} +
    h^{a}{}_{\rho} \, \partial_{\nu} h_{b}{}^{\rho} = 0. 
    \label{eq:gambartombar} 
\end{equation} %

    \item Justifying the name
``teleparallelism'', it parallel-transports each vector of the tetrad
$\{h_{a}\}$ everywhere:
    \begin{equation}
	 {\bar \nabla}_{\lambda} \ h_{a}{}^{\mu} \equiv \partial_{\lambda} \
	 h_{a}{}^{\mu} + {\bar \Gamma}^{\mu}{}_{\rho \lambda}
	 h_{a}{}^{\rho} = 0.
        \label{eq:paratransp}
    \end{equation}
    
    \item In consequence, it preserves the metric $g$:
${\bar \nabla}_{\lambda} \ g_{\mu \nu} = 0$. \vspace{-6pt}

\item It has vanishing Riemann curvature tensor:
${\bar R}^{\rho}{}_{\sigma \mu
\nu} = 0$. \vspace{-6pt}

\item It has a non-vanishing torsion ${\bar T}$:
\begin{equation}
    {\bar T}^{\lambda}{}_{\nu \mu} = h_{a}{}^{\lambda} \left(
    \partial_{\nu} h^{a}{}_{\mu} - \partial_{\mu} h^{a}{}_{\nu}
    \right) = h_{c}{}^{\lambda} f^{c}{}_{a b}\  h^{a}{}_{\mu}
    h^{b}{}_{\nu}.
    \label{eq:barT}
\end{equation}
In the frame $\{h_{a}\}$ itself, this torsion is pure anholonomy and,
consequently, a measure of the non-triviality of the metric $g$. \vspace{-6pt}

\item The Levi-Civita and the Weitzenb\"ock connections are related by
\begin{equation}
\gammabol_{\lambda \mu \nu} = {\bar \Gamma}_{\lambda (\mu \nu)} - 
{\bar T}_{[\mu \nu] \lambda }.
\end{equation}

\item  In consequence the geodesic equation of General Relativity acquires,
in terms of ${\bar \Gamma}$, the form of a force equation:
\begin{equation}
\frac{d U^{\lambda}}{du} + {\bar \Gamma}^{\lambda}{}_{\mu \nu}\
U^{\mu} U^{\nu} = -\ {\bar K}^{\lambda}{}_{\mu \nu}\ U^{\mu} U^{\nu}.
    \label{eq:geodesic5}
\end{equation}

\item  Use of (\ref{decomp1}) for the specific case of the 
Weitzen\-b\"ock connection gives its contorsion as 
\begin{equation}
{\bar K}^{\lambda}{}_{\mu \nu} = h_a{}^{\lambda}\ \omegabol^a{}_{b
\nu}\ h^b{}_{\mu}.
\label{contorsion4}
\end{equation}
This means that $\omegabol$ is the Weitzenb\"ock contorsion~\cite{agp1} seen from
the frame $\{h_a\}$:
\[
\omegabol^a{}_{b c} =
{{\stackrel{\circ}{\bar K}}{}}{^a{}_{b c}}.
\] 

\item The index symmetries give to
the force equation (\ref{eq:geodesic5}) the form
\begin{equation}
\frac{d U^{\lambda}}{du} + {\bar \Gamma}^{\lambda}{}_{\mu \nu}\
U^{\mu} U^{\nu} =
h_a{}^{\lambda} f_b{}^a{}_c\ U^{b} U^{c}.
    \label{eq:geodesic6}
\end{equation}
The right-hand side ``force'' is one more measure of the tetrad
non-holonomy.
 \end{enumerate}

The Weitzen\-b\"ock connection is a kind of (curvature) ``vacuum'' of every other
connection. In fact, a general connection with holonomic
components $\Gamma^\lambda{}_{\mu \nu}$ will be related to its non-holonomic
components $\omega^a{}_{b \nu}$ by
\[
\Gamma^\lambda{}_{\mu \nu} = h_a{}^\lambda \, \omega^a{}_{b \nu} \,
h^b{}_\mu + {\bar \Gamma}^{\lambda}{}_{\mu \nu},
\]
which is actually the inverse of (\ref{eq:omegagain}), with a further
substitution of Eq.~(\ref{eq:weitzen}). Suppose then we look at the
Weitzen\-b\"ock connection of a tetrad $h^{a}{}_{\mu}$ from another tetrad
$h^{'a}{}_{\mu}$. It will have the expression
\begin{equation}
 {\bar \omega}^{' a}{}_{b \nu} = h^{' a}{}_{\lambda} {\bar
 \Gamma}^{\lambda}{}_{\mu \nu}\ h'_{b}{}^\mu + h^{' a}{}_{\mu}
 \partial_\nu h'_{b}{}^\mu = (\Lambda\ \partial_\nu
 \Lambda^{-1})^a{}_b,
\end{equation}
with $\Lambda$ as given by Eq.~(\ref{Lortetrad}). The Weitzen\-b\"ock
connection of a tetrad $h^{a}$, when looked at from another tetrad $h^{'
a}$, is the vacuum of a gauge theory for the Lorentz group (whose corresponding
field strength would be the curvature tensor). The ``gauge'', or the group
element, is just that relating the two connections.

We can also consider the difference between the Weitzen\-b\"ock
connections of two different tetrad fields.  If ${\bar \Gamma}$ is
related to $h^{a}$ and ${\bar{\Gamma}}'$ to $h^{' a}$, then
\begin{equation}
{\bar \Gamma}^{\lambda}{}_{\mu \nu} - {\bar{
\Gamma}}'{}^{\lambda}{}_{\mu \nu} = h^{'}{}_a{}^{\lambda}\ h^{' b}{}_\mu
\ [\Lambda \partial_\nu \Lambda^{-1}]^a{}_b, \label{twoWeitz}
\end{equation}
which tells us that two distinct tetrads, $h_a$ and $h^{'}{}_a$, can have the same
Weitzen\-b\"ock connection. In that case, they differ by a point-independent
Lorentz transformation. Along a curve of parameter $u$, the  accelerations defined
by these connections will differ as
\begin{equation}
{\bar a}^{\lambda} - {\bar {a}}{}'^{\lambda} =
h^{'}{}_a{}^{\lambda}\ h^{' b}{}_\mu \ U^\mu\ U^\nu\ \left[\Lambda  \partial_\nu
\Lambda^{-1} \right]^a{}_b.
\end{equation}

\section{Final Comments}

There is a functional six-fold infinity of tetrad fields determining a
given metric as in Eq.~(\ref{eq:tettomet}). This six-foldedness is
``functional'' because such tetrad fields differ by point-dependent
(that is, local) Lorentz transformations.  Anholonomy is essential to
the presence of a gravitational field: All holonomic tetrads
correspond to  Minkowski flat space.  Each tetrad field defines
also a Weitzen\-b\"ock flat connection, whose torsion measures
its anholonomy and represents, in the teleparallel approach, the
gravitational field strength.  There is a (non-functional) six-fold
infinity of tetrad fields with the same Weitzen\-b\"ock
connection, differing from each other by point-independent (global)
Lorentz transformations.  As each result of \GR\ can be stated in terms
of the tetrad anholonomy, gravitation  reduces to frame effects. 
In \GR\ the absence or presence of gravitation is signaled by the vanishing or not
of a {\em covariant} derivative, the curvature tensor.  The field is
a ``covariant'' anholonomy. In teleparallelism, the presence of field is signaled
by a {\em simple} anholonomy, that of the tetrad field itself. In the tetrad
frame, everything happens in Minkowski space, but the frame will be, we insist,
necessarily anholonomic.  A holonomic velocity in Riemann space becomes,
once written with components in  the tetrad frame, an anholonomic
velocity in flat tangent Minkowski space.

A better understanding of the relationship between the standard
formulation of \GR\ and teleparallelism is still necessary. In
particular, it should be decided which field is fundamental --- metric
or tetrad. The equivalence of both approaches may come to disappear
at the quantum level. If an interaction is mediated by a spin-2
field, matter can attract both matter and antimatter, but mediating
vector (spin-1) fields would give opposite signs for matter-matter and
matter-antimatter interactions~\cite{kibble}. Antimatter produced by
high-energy matter collisions, however small its amount, would produce a
cosmic repulsion. Whether or not the exchange of constrained four-vectors can
be equivalent to that of a spin-2 field is an open question.

\begin{acknowledgments}
The authors thank FAPESP-Brazil and CNPq-Brazil for financial support.
\end{acknowledgments}

\end{document}